# Unravelling compound risks of hydrological extremes in a changing climate: Typology, methods and futures

Kwok P Chun[1], Thanti Octavianti[2], Georgia Papacharalampous[3], Hristos Tyralis[4], Samuel J. Sutanto[5], Pavel Terskii[6], Paola Mazzoglio[7], Dario Treppiedi[8], Juan Rivera[9], Nilay Dogulu[10], Adeyemi Olusola[11], Bastien Dieppois[12], Moctar Dembélé[13], Simon Moulds[14], Cheng Li[15], Luis Alejandro Morales-Marin[16], Neil Macdonald[17], Toundji Olivier Amoussou[18], Roland Yonaba[19], Salomon Obahoundje[20], Nicolas Massei[21], David M. Hannah[22], Sivarama Krishna Reddy Chidepudi[23], Byman Hamududu[24]

[1] School of Architecture and Environment, University of the West of England, kwok.chun@uwe.ac.uk https://orcid.org/0000-0001-9873-6240
[2] School of Architecture and Environment, University of the West of England, thanti.octavianti@uwe.ac.uk https://orcid.org/0000-0002-9921-0539
[3] Department of Topography, School of Rural, Surveying and Geoinformatics Engineering, National Technical University of Athens, Greece papacharalampous.georgia@gmail.com https://orcid.org/0000-0001-5446-954X
[4] Construction Agency, Hellenic Air Force, Greece, montchrister@gmail.com https://orcid.org/0000-0002-8932-4997
[5] Water System and Global Change group, Wageningen University and Research, samuel.sutanto@wur.nl https://orcid.org/0000-0003-4903-6445
[6] Hydrology research unit at Swedish Meteorological and Hydrological Institute, Sweden; Water Problems Institute of the Russian Academy of Sciences, Moscow, Russia; skipper.k2000@gmail.com https://orcid.org/0000-0001-7852-8241
[7] Department of Environment, Land and Infrastructure Engineering (DIATI), Politecnico di Torino, Italy, paola.mazzoglio@polito.it, https://orcid.org/0000-0002-3662-9439
[8] Dipartimento di Ingegneria, Università degli Studi di Palermo, Italy, dario.treppiedi@unipa.it, https://orcid.org/0000-0002-1359-1650
[9] Instituto Argentino de Nivología, Glaciología y Ciencias Ambientales (IANIGLA), CCT-Mendoza/Consejo Nacional de Investigaciones Científicas y Técnicas (CONICET), Mendoza, Argentina, jrivera@mendoza-conicet.gob.ar, https://orcid.org/0000-0001-7754-1612
[10] World Meteorological Organization, Genève, Switzerland nilay.dogulu@gmail.com , https://orcid.org/0000-0003-4229-2788
[11] Faculty of Environmental and Urban Change, Toronto, Canada aolusola@yorku.ca https://orcid.org/0000-0003-2295-5214
[12] Centre for Agroecology, Water and Resilience (CAWR), Coventry University, Coventry, UK ab9482@coventry.ac.uk https://orcid.org/0000-0001-7052-1483
[13] International Water Management Institute (IWMI), Accra, Ghana, moctar.dembele@cgiar.org, https://orcid.org/0000-0002-0689-2033
[14] University of Edinburgh, smoulds@ed.ac.uk https://orcid.org/0000-0002-7297-482X
[15] Department of Ecology, Yangzhou University, China, licheng@yzu.edu.cn
[16] The National University of Colombia in Bogota lmoralesm@unal.edu.co
[17] Department of Geography and Planning, University of Liverpool, Neil.Macdonald@liverpool.ac.uk https://orcid.org/0000-0003-0350-7096

[18] Higher Institute of Sustainable Development, University of Fada N'Gourma, Fada N'Gourma, Burkina Faso, toundji.amoussou@univ-fada.bf, https://orcid.org/0000-0002-8538-2305

[19] Hydro-systems and Agriculture (LEHSA), International Institute for Water and Environmental Engineering (2iE), Burkina Faso, ousmane.yonaba@2ie-edu.org https://orcid.org/0000-0002-3835-9559

[20] International Joint Laboratory (LMI-NEXUS), Université Félix Houphouët Boigny, Abidjan, Côte d'Ivoire; International Water Management Institute (IWMI), Accra, Ghana. Obahoundjes@yahoo.com , https://orcid.org/0000-0001-8093-5241

[21] University of Rouen Normandie, France, nicolas.massei@univ-rouen.fr, https://orcid.org/0000-0001-9951-4041

[22] School of Geography, Earth & Environmental Sciences, University of Birmingham, d.m.hannah@bham.ac.uk https://orcid.org/0000-0003-1714-1240

[23] University of Rouen Normandie, France, sivaramakrishnareddy.chidepudi@univ-rouen.fr https://orcid.org/0000-0001-9394-7970

[24] Norwegian Water Resources and Energy Directorate, bymanh@gmail.com byha@nve.no

**Abstract:** We have witnessed and experienced increasing compound extreme events resulting from simultaneous or sequential occurrence of multiple events in a changing climate. In addition to a growing demand for a clearer explanation of compound risks from a hydrological perspective, there has been a lack of attention paid to socioeconomic factors driving and impacted by these risks. Through a critical review and co-production approaches, we identified four types of compound hydrological events based on autocorrelated, multivariate, and spatiotemporal patterns. A framework to quantify compound risks based on conditional probability is offered, including an argument on the potential use of generative Artificial Intelligence (AI) algorithms for identifying emerging trends and patterns for climate change. Insights for practices are discussed, highlighting the implications for disaster risk reduction and knowledge co-production. Our argument centres on the importance of meaningfully considering the socioeconomic contexts in which compound risks may have impacts, and the need for interdisciplinary collaboration to effectively translate climate science to climate actions.

**Keywords:** Compound risks, hydrology, extreme events, interdisciplinary, co-production, conditional probability

# INTRODUCTION

In recent decades, the intensity, frequency, and duration of climate and hydrological extremes, including compound events, have increased across many regions of the world. Compound events (including risks, extremes and hazards) refer to the simultaneous or sequential occurrence of multiple risks, which can interact, exacerbate, or create cascading effects. For example, paradoxical events like floods in California during a multi-year drought in 2023 challenge existing understanding about large-scale variation effects on hydrological extremes. These emerging joint extreme conditions are believed to be linked to large-scale atmospheric variations and teleconnections, fuelled by global warming due to increased greenhouse gas emissions. This calls for innovative approaches for its characterisation and novel actions to mitigate its impacts. Projected growth in variability of extreme weather events may consequently cause disastrous combination of drivers (IPCC, 2023) further exacerbated by anthropogenic influences.

These impacts and future projections illustrate the vulnerability of societies to hydro-meteo-geological extremes and demonstrate the need for improved understanding, forecasting and prediction of such events individually (Blöschl et al., 2013), but also as compound (Hillier and Dixon, 2020) and concurrent events (Hillier et al., 2015). While various hydrological events and extremes have been studied and described in hundreds of publications, a unified explanation of compound hydrological risks, extremes or events, their interconnections and relations to cascading, joint and concurrent extremes of various spatial, temporal and relational scales is lacking. Traditional methods of risk assessment often focus on single events, but this can be misleading in the context of compound risk where the assumption of independence is violated.

Several factors are missing in the current studies and required special attention. The characteristics of compound events associated with natural phenomena are expected to change due to climate change. Future scenarios with more frequent hydroclimatic extremes may be more likely than previously projected. New approaches are needed to account for the joint probability of multiple extreme events, providing a clearer explanation of how compound risk evolves over time based on nonstationary statistical theory, and how it is shaped by climate change.

Furthermore, there is a notable disconnection between progress in climate science and their practical application. Socioeconomic factors can amplify hazards and their impacts on society, particularly for marginalised communities that have historically borne the brunt of disasters. Conversely, resilient social systems including informed communities, robust infrastructures, both proactive and reactive strategies can greatly reduce the impacts of compound events and accelerate recovery. This interaction between natural processes and socioeconomic systems should be meaningfully considered when undertaking investigations on compound hydrological extremes. The complexity of our networked society calls for a better understanding of interconnected risks, whereas traditional statistical models may not capture all interconnections. Therefore, nuanced analyses that account for diverse factors are needed.

In this perspective, we delve into the compound risk definitions and methods of synthesis and typology, learning through co-production approaches fostered by interdisciplinary discussions. Our novel approach lies in the concept of Compound Typology, which is a unique blend of climate, hydrology, human interaction, and environmental factors. To develop our argument, we (i) explore current studies in the field to identify a typology of compound hydrological events; (ii) propose methods, focusing on conditional distributions and extreme value theory, and (iii) discuss insights for practices and identify future research directions on this field.

**A TYPOLOGY OF COMPOUND HYDROLOGICAL EVENTS**

Compound hydrological risk is a relatively new concept that refers to the risk of multiple extreme hydrological events occurring simultaneously or in close succession in a changing climate. Quantifying and managing this type of risk can be challenging due to the unpredictability of interactions between various extreme events (de Ruiter et al., 2020).

The definition of compound risk for climate change in this study is derived from diverse perspectives and their corresponding counter arguments, which have been co-produced by 25 EURO-FRIEND researchers[1]. Using critical review and co-production approaches, we started by identifying more than 600 papers and reviewed a total of 277 relevant papers. The post-review process involved researchers independently providing their definition of hydrological compound events. Further details on the interactive review process can be found in the supplementary material.

A Compound Hydrological Event (CHE) manifests as the simultaneous or sequential occurrence of multiple (non) hydrological hazards, either in the same or neighbouring catchments. While the hazards can be hydrological, such as floods and hydrological droughts, or non-hydrological, such as heatwaves, fires, and landslides, at least one hydrological hazard must be present. A cascading hydrological event represents a subtype of CHE where hazards occur sequentially within a short time frame. It is worth noting that there is no clear guidance on the time gap between one hazard to another to determine if the events have cascading relationship. Based on this definition, we further categorise the reviewed papers into CHE scenarios, with either hydrological event serves as the primary or secondary hazard and whether the event entails co-occurrence and/or sequential (Table 1).

Table 1. Examples of compound event studies categorized into hydrological and non-hydrological hazards as primary and secondary hazards, respectively, and vice versa

| Event scenario | Primary event | Secondary event | Nature of event | Examples | References |
|---|---|---|---|---|---|
| Hydrological event | Hydrological event | - | Co-occurrence | 1. Compound inland flood with high downstream water level<br>2. Compound flood downstream due to flooding in tributaries<br>3. Compound flood in different regions in the US | 1. Wang and Shen (2023), Ma et al. (2021)<br>2. Najibi et al. (2023) |
| | | | Sequential | 1. Flood in Germany 2016 followed by flash flood in the next days and in another location due to clogging and sediment load<br>2. Cascading flood and streamflow drought within 6-month durations<br>3. Cascading flood and outburst flood<br>4. Flood event followed by another flood in a month | 1. Thieken et al. (2022)<br>2. Rezvani et al. (2023)<br>3. Fischer et al. (2023)<br>4. Knighton and Nanson (2001) |
| Hydrological and non-hydrological | Hydrological event | Non-hydrological event | Co-occurrence | 1. Compound extreme river discharge with storm surge or high tide<br>2. Compound fluvial flooding and sea level rise<br>3. Compound coastal flood due to high groundwater table, sea level rise, and coastal precipitation | 1. Heinrich et al. (2023), Zhang and Chen (2022), Camus et al. (2022), Zhang et al. (2020), Herdman et al. (2018)<br>2. Moftakhari et al. (2017), Ghanbari et al. (2021), Del-Rosal- |

| | | | | | 3. Salido et al. (2021) |
| | | | | | 4. Rahimi et al. (2020), Peña et al. (2023) |
| | | | Sequential | 1. Compound and cascading events of soil moisture drought, heatwave, and fires in Europe<br>2. Sequential flood and heatwave events<br>3. Cascading flood and landslide events | 1. Sutanto et al. (2019)<br>2. Chen et al. (2021), Liao et al. (2021), Gu et al. (2022)<br>3. Chen et al. (2023) |
| | Non-hydrological event | Hydrological event | Co-occurrence | 1. Compound hot extreme and soil moisture drought and/or runoff drought<br>2. Compound heat stress and flooding<br>3. Compound meteorological and hydrological droughts<br>4. Compound El Niño (rising temperature) and precipitation, runoff, and floods<br>5. Compound effect of rainfall and snowmelt on flood events in one place | 1. Hao et al. (2018), Feng et al. (2022, 2023)<br>2. Zhang and Villarini (2020)<br>3. Wu et al. (2022), Sarhadi et al. (2023)<br>4. Liu et al. (2018)<br>5. Sezen et al. (2020) |
| | | | Sequential | 1. Heavy rainfall in the upstream followed by floods and/or landslides at the downstream regions<br>2. Cascading events of earthquake, landslide, debris flow, and flooding<br>3. Cascading meteorological and hydrological droughts<br>4. Cascading landslide, | 1. Schauwecker et al. (2019), Talchabhadel et al. (2023), Biondi et al. (2023)<br>2. Zhang et al. (2023)<br>3. Vorobevskii et al. (2022)<br>4. Geertsema, et al. (2022) |

| | | | | tsunami, and flood | |
|---|---|---|---|---|---|

CHE with hydrological events as the primary event is typically instigated by a single hazard, occurring at different locations and/or time. For instance, high river discharges in upstream river tributaries lead to or amplify floods in downstream areas (Ma et al., 2021; Wang and Shen, 2023). Sequential CHE may involve not only the same hazard but also different hazards, such as hydrological drought.

In cases of co-occurrence between hydrological and non-hydrological events, separating the primary and secondary hazards is less straightforward. In our example, floods are classified as the primary hazard in the compound of coastal floods alongside non-hydrological hazards, such as sea level rise, storm surge, or high tide (e.g., Moftakhari et al., 2017; Rahimi et al., 2020; Heinrich et al., 2023) since floods constitute the major hazard. However, many studies on the sequentially of CHE provide explicit examples, distinguishing between primary and secondary hazards. For example, flood event followed by hot extremes or heatwaves several months later, or flood events followed by landslides a few days later, serve as clear examples where hydrological hazards act as the primary hazard (Chen et al., 2021; Gu et al., 2022; Chen et al., 2023).

When a CHE features a non-hydrological event as the primary event, a clear relationship between primary and secondary hazards often emerges. Many studies analysing the co-occurrence of non-hydrological hazards with hydrological hazards identified meteorological or climatic hazards as the primary hazards. This is due to numerous hydrological events are triggered by meteorological hazards e.g., high temperature and high or low precipitation. Precipitation and temperature are frequently perceived as drivers rather than hazards, which play a crucial role in the occurrence of floods and droughts. Nonetheless, there is ambiguity regarding the definition whether heavy rainfall leading to floods should be considered as a compound event or simply a natural occurrence. For hydrological considerations, we exclusively consider precipitation events that trigger a hazard followed by another hazard as CHE. For instance, low precipitation and high temperature result in meteorological drought, subsequently propagating into hydrological drought (Wu et al., 2022; Sarhadi et al., 2023). In this scenario, low precipitation and high temperature are classified as drivers for meteorological drought.

Based on this review, we identified four types of compound hydrological events:

1. Emerging Preconditioned Events: These occur when weather or climate conditions exacerbate the effects of a hazard (e.g., heavy precipitation triggers flood and then landslide).

2. Temporally Compounding Events: These events result from a sequence of hazards unfolding over time (e.g., cascading flood and streamflow drought within a 6-month period).

3. Spatially Compounding Events: Hazards occurring in interconnected locations collectively contribute to an aggregated impact (e.g., flood in the west and east coasts of the US or upstream and downstream relation).

4. New Combination of Multivariate Events: These involve multiple drivers or hazards converging to produce an impact (e.g., effect of the El Niño-Southern Oscillation on precipitation and runoff leading to floods in Texas).

The four types described above are all influenced by climate change. In the emerging preconditioned events, changes in regional circulations alter hydroclimatic patterns, affecting the severity and frequency of related hazards. Changing climate has played a role in shifting these regional circulations and therefore these types of hazards are likely to get worse in the future (IPCC, 2023). This is also the case for the temporally and spatially compounding events. Time and scale are key dimensions characterising non-stationarity and spatially heterogenous impacts of climate change have been observed as a result of compound events (Zscheischler et al. 2020). The new combination of multivariate events is influenced by global and regional climate changes, and this leads to unprecedented hydroclimate-related risks. Uncertainties, sourced from model uncertainties, scenario uncertainties and internal variability, features very strongly in these known of unknown events.

The manifestations and implications of these four types of compound events to our everyday life are identified. For the emerging preconditioned events, improving societal resilience through building robust and adaptive infrastructural, information and institutional systems becomes key in events exacerbated by a weather or climate condition. The temporally compounding events would require responding to time-sensitive events and understanding their potential to amplify or create another hazard, considering the time needed to recover before another event hit (de Ruiter et al., 2020). The spatially compounding events concerns about the scale spatiality and responsibility of authorities. As events may occur at varying scale and do not abide by any political boundaries, it is important to ensure that authorities are clear of their responsibility of managing compound risks. Rethinking the distribution of resources of any types including funding, projection capacity, emergency services, etc, is essential. Finally, the multivariate events require thinking about worst case scenarios, creating plans about unprecedented events, developing new approach in urban planning and building capacity with limited resources and residual risks that may not be absorbed by current mitigation measures.

Overall, compound events for hydrological risk management and climate resilience are the intricate interplay of environmental factors and their complex interaction over various spatial and temporal scales. The typology discussed above, along with insights from Zscheischler et al. (2020), can be applied to modelling algorithms for compound risk, where all components are treated as random variables.

**QUANTIFYING AND UNDERSTANDING COMPOUND RISK IN A COMMUNITY**

The properties of extreme events are defined by second-order moments related to variances and cross-correlations, as well as higher-order moments, such as the third moment, for the tail extreme distributions. Climate change challenges the traditional assumptions of stationary and ergodicity in compound risk analysis, which have primarily focused on mean change. This makes interactions between extraordinary events difficult to predict directly. Emerging compound events, driven by new environmental extremes, result in the coupling of distinct natural and social phenomena. At first glance, when two independent extreme events occur, their compound probabilities can be simply multiplied together. However, when new extreme phenomena are synchronised and coupled due to climate change, the compound risk of extreme events can increase significantly.

The term "event" necessitates statistical modelling. As adapted from Zscheischler et al. (2020) and synthesized from the compound risk literature of the previous section, Figure 1 outlines a model for compound events. Specifically, a compound event occurs when at least two drivers or hazards materialise. The output of such a model, which is also of interest from a socioeconomic perspective, is frequently the hazards.

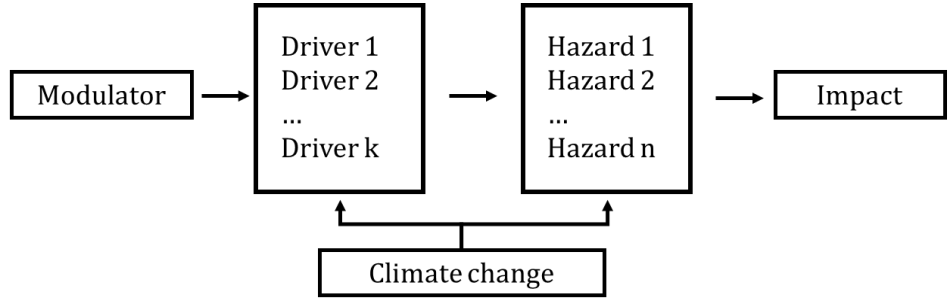

Figure 1. Modelling procedure of compound events.

Statistical modelling suggests that hazards should be modelled as random variables, thereby assigning probability distributions to them. In the absence of information on modulators, drivers and climate change, hazards can be modelled directly with a joint probability distribution $P(H_1 \leq h_1, \ldots, H_n \leq h_n)$ where the multidimensional random variable $\boldsymbol{H} = (H_1, \ldots, H_n)$ denotes multivariate hazards (case 1). A popular modelling framework, for such cases are copulas.

But when materializations $\boldsymbol{m}$ of the modulator, $\boldsymbol{d} = (d_1, \ldots, d_k)$ of the drivers and $\boldsymbol{c}$ of climate change variables are known, the conditional predictive joint probability distribution of hazards $P(H_1 \leq h_1, \ldots, H_n \leq h_n | \boldsymbol{d}, \boldsymbol{m}, \boldsymbol{c})$ must be estimated (case 2). The concept of driver has been discussed in the previous section. When talking about climate change materialization $\boldsymbol{c}$, we might refer to Global Circulation Models (GCMs) or Regional Circulation Models (RCMs) data. Modulators are weather patterns (e.g. an El Niño event) that are causally related to drivers.

In the following, without loss of generality, hazards are modelled as continuous variables. Consequently, the problem is restricted to regression and there are two options to model/predict the hazards:

a. Predicting the conditional cumulative distribution function (CDF) $F$ of the hazard, of the $M_F$ distributional regression model,

$$F(\boldsymbol{h}|\boldsymbol{d}, \boldsymbol{m}, \boldsymbol{c}) = M_F(\boldsymbol{d}, \boldsymbol{m}, \boldsymbol{c}, \boldsymbol{\theta}), \qquad (1)$$

The most well-known examples are generalized additive models for location, scale and shape (GAMLSS) (Rigby and Stasinopoulos 2005).

b. Predicting a possibly multidimensional statistical functional $T$ (e.g. the conditional mean (quantile), or a multidimensional mean (quantile)) of the conditional probability distribution of the $M_T$ semi-parametric model (Dimitriadis et al. 2024)

$$T(F(\boldsymbol{h}|\boldsymbol{d},\boldsymbol{m},\boldsymbol{c})) = M_T(\boldsymbol{d},\boldsymbol{m},\boldsymbol{c},\boldsymbol{\theta}), \quad (2)$$

The most-well known examples of semi-parametric models are least-squares linear regression and quantile regression (Koenker and Bassett, Jr 1978).

In both cases, the parameters $\boldsymbol{\theta}$ of the models must be estimated, using (a) proper scoring rules for the case of distributional regression (Gneiting and Raftery 2007), or (b) consistent scoring functions in the case of semi-parametric models (Dimitriadis et al. 2024).

Equations (1) and (2) allow modelling the four identified types of compound events, where the selection of the type of model (e.g. time series model or spatial model) should be guided by the typology of the compound event. For instance, for the case of a single hazard ($n = 1$) and multiple drivers that evolve in time, $t = 1, ..., u$ in a single location $s = 1$, a time series model (Box et al. 2015) or a related deep learning model (e.g. LSTM networks, Hochreiter and Schmidhuber 1997) would be more appropriate. In the case of drivers at multiple locations $s = 1, ..., v$, at time $t = 1$, Kriging (see Cressie 1990 and references therein) or a convolutional neural network (CNN, Le Cun et al. 1989) has been widely adopted.

Regardless of the model type (temporal, spatial or spatiotemporal), the scope is that its predictions should be probabilistic (Gneiting and Raftery 2007) or expressed in terms of multiple functionals (Hallin et al. 2010) of various types (Newey and Powell 1987), potentially focusing on regions of extremes (Allouche et al. 2024; Daouia, et al. 2018; Lerch et al. 2017; Taggart 2022; Wang et al. 2012). Relevant advances in statistical modelling and AI that might interest the climate community are summarised by Papacharalampous and Tyralis (2022) and Tyralis and Papacharalampous (2024).

Another challenge in modelling compound events lies in determining the associated hazards, drivers, and distinguishing between primary and secondary hazards. In the context of the proposed formula, $H_n$ can represent either hydrological or non-hydrological hazards (see Table 1). For instance, in the case of the two-dimensional hazard of heatwaves (a meteorological hazard) and soil moisture drought, heatwaves can accelerate soil drying, leading to drought, but conversely, extreme drought conditions can exacerbate high temperatures (Rasmijn et al., 2018; Teuling, 2018). Furthermore, the distinction between hazards and drivers is often ambiguous. Many studies classify the occurrence of heavy precipitation and subsequent flooding as compound hazards ($H_1$ and $H_2$ see Table 1). However, heavy precipitation can also be viewed as a driver ($d_1$) of flooding. Similarly, a few consecutive days of high temperatures may be identified as heatwaves ($H_1$), yet these conditions can also act as a driver ($d_1$) for drought. To address this, we propose that the definitions of hazards and drivers should be clarified. For example, rainfall below a certain threshold may be considered a driver, but when it exceeds that threshold, it should be categorised as a hazard.

A new AI approach paradigm based on generative algorithms is needed to identify emerging trends and patterns to understand how physical model outputs and observations could be used to constrain unprecedented but possible compound risk events for new emerging behaviours in earth systems based on mass, energy, and momentum conservations. AI can learn from ensembles based on ergodicity assumptions for extreme events. The results can be made more transparent by using statistical downscaling techniques. To train AI, it is critical to keep high-order moment information in AI outputs when probable extremes are contained in ensembles of outputs from numerical models in atmospheric science, hydrology, and other environmental disciplines.

Furthermore, theory development is critical for the compound risk assessment and conceptualisation from observation and model outputs. Theoretically, the extreme distribution should be independent, and the joint extreme distribution based on the law of large numbers is simply the product of their marginal distributions of individual extremes. The interesting future aspect is that the overlap or correlations of extremes cause the scaling relationship, and therefore, there is a need for new statistical theory to explain the emergence of compound risks based on statistical joint relationships and independence.

What is fundamentally missing from the current approach is the consideration of socioeconomic factors, such as demographics, housing conditions, access to resources, social networks, institutional capacity, infrastructural readiness, among others. In Figure 1, hazard prediction is the focus, while impacts associated with socioeconomic evaluations are overlooked. These factors can exaggerate the impact of these compound events by weakening the social fabric necessary for adaptative action. Methods of evaluating compound risks should incorporate not only physical environmental data but also social data. Both narrow and broad approaches to quantifying compound risks have their own merits; however, we advocate for a broader perspective that includes socioeconomic factors. These approach enables a more meaningful understanding of how compound risks in a changing climate can be effectively addressed within a community.

**CONCLUSION**

Our review suggests that in the realm of climate change, the newly hydrological hazard patterns which are autocorrelated, multivariate, and spatiotemporally patterned are observed. In the short term, these patterns need new algorithms to model and predict to quantify emerging variations related to climate change. Whereas in a longer term, an integrative way of managing risks should be implemented. The management of these new phenomena is a complex task, fraught with barriers and problematic concepts. Two main points we aim to emphasise here are the importance of considering socioeconomic factors when studying compound risks and fostering interdisciplinary collaboration in this effort.

Firstly, the absence of socioeconomic factors from the assessment and quantification of compound events risks may exacerbate their impacts on vulnerable communities. Therefore, it is crucial to broaden the focus by explicitly considering the interactions between natural processes and socioeconomic systems for climate change. Compound events have caused disproportionate loss and damage to vulnerable communities, such as the Idai and Kenneth cyclones in Mozambique (Norton et al., 2020) killing approximately 1303 and affected more than three million people. Some areas are inherently prone to compound risks due to its topography, geography and other natural factors, such as in Sub Sahara Africa and Small Island Developing States (e.g. Bush 2018, Forbes-Genade and Chenga, 2019). This issue is further exacerbated by the lack of resources and capacity to prepare for, mitigate and adapt to such events. When viewed through the lens of the four types of compound events, this challenge manifests in various ways: slow or inadequate recovery from disasters in preconditioned events; difficulties in resource distribution and coordination pre-during-post events in spatio-temporally patterned events; and challenges in understanding and preparing for multivariate events due to projection of unknown risks. Consequently, international collaboration becomes essential for reducing the impacts of compound risks and protecting all communities, particularly those that are most vulnerable.

Our co-production approach in writing this perspective reflects our commitment to the growing call for collaboration. A place-based understanding is particularly important in the global majority regions, where the impacts of climate change can be unique and severe. We not only considered the local context but also integrated international research interactions, which are essential for generating knowledge related to climate actions for compound risks. This co-production process is instrumental in challenging the colonial power structures within international initiatives. Even among hydroclimate scientists, there is a recognised need for greater inclusivity in defining and analysing compound risks, particularly from perspectives that may be difficult to engage with. Whilst this approach has its shortcomings, including the significant time required for meetings and logistics, it is a worthwhile effort to foster shared ownership of knowledge.

Secondly, training within a single discipline often shapes a specific worldview, potentially limiting researchers' perspective and leaving critical blind spots unaddressed. This challenge can only be adequately met through an interdisciplinary approach. Studying emerging compound risks supports the implementation of the Sendai Framework for Disaster Risk Reduction (SFDRR) (United Nations Office for Disaster Risk Reduction, 2015), and several new approaches are being developed to manage compound risks. A promising approach is the use of integrated risk management frameworks (e.g. Giupponi et al., 2015), which bring together various stakeholders, including governments, academics, businesses, and communities, to develop coordinated responses to compound risks. An approach that promotes and encourages interdisciplinary and intersectoral collaboration is required for creating impactful climate actions. Strategies can include diversifying critical infrastructures, developing early warning systems for multi-hazards, integrating aquatic or semi-aquatic animal behaviour into warning systems and developing contingency plans. Emerging initiatives, such as the Early Warnings for All (EW4All), which officially became the World Meteorological Organization (WMO)'s top priority since 31 May 2023, are examples on this shift.

In conclusion, it is necessary to break traditional thought patterns within the scientific community to embrace interdisciplinary approaches for climate action and use AI to address complex environmental issues. We believe in the power of co-production, shared knowledge, and the importance of responding to and learning from compound events. We hope this perspective will inform strategies to address climate change and contribute to the emerging interdisciplinary work aimed at bridging the gap between progress in climate science and practices on the ground.

**Notes:** EURO_FRIEND is a regional group focused on hydrology, established in 1985. It includes 31 European countries and operates under the UNESCO Centre for Water Resources and Global Change (ICWRGC). The group aims to bridge data and knowledge gaps in hydrology by sharing a comprehensive hydrological database. This paper is a result of the cross-sectional Theme C group of EURO-FRIEND, led by the University of the West of England, established in 2023 during the European Geoscience Assembly Splinter meeting.

**Conflict of interest:** The authors declare that there are no conflicts of interest related to this publication.